\documentclass[10pt,final,journal,journal,comsoc]{IEEEtran}

\usepackage{graphicx,cite,epsfig,amsmath,amsfonts,multicol,subfigure,mathtools,bm,mathrsfs,setspace,algorithm,algorithmic,amsthm,dblfloatfix}
\usepackage{amssymb}
\usepackage{amsmath}
\usepackage{graphicx}
\usepackage{epstopdf}
\usepackage{multirow}
\usepackage{color}
\usepackage{enumitem}
\usepackage{cite}
\theoremstyle{plain}
\newtheorem{lemma}{Lemma}
\begin{document}
	\title{Dynamic NOMA-Based Computation Offloading in Vehicular Platoons} 
	\author{Dongsheng Zheng, 
		Yingyang Chen, \IEEEmembership{Member,~IEEE},
		Lai Wei,
		Bingli Jiao, \IEEEmembership{Senior Member,~IEEE},
		and Lajos Hanzo, \IEEEmembership{Life Fellow}
	\thanks{This work was supported in part by the National Natural Science Foundation of China under Grant 62001191 and 62171006, in part by the National Key Research and Development Program of China under Grant 2020YFB1807802, in part by the Guangdong Basic and Applied Basic Research Foundation under Grant 2023A1515012892 and 2021B1515120067, and in part by the Guangzhou Science and Technology Project under Grant 202201010256. L. Hanzo would like to acknowledge the financial support of the Engineering and Physical Sciences Research Council projects EP/W016605/1 and EP/X01228X/1 as well as of the European Research Council's Advanced Fellow Grant QuantCom (Grant No. 789028) (\emph{Corresponding author: Yingyang Chen}.)}
		\thanks{D. Zheng, L. Wei, and B. Jiao are with the Department of Electronics, Peking University, Beijing 100871, China (e-mail:
			zhengds@pku.edu.cn; future1997@pku.edu.cn; jiaobl@pku.edu.cn).}
		\thanks{Y. Chen is with the College of Information Science and Technology, Jinan University, Guangzhou 510632, China, and also with the Guangdong Provincial Key Laboratory of Data Security and Privacy Protection, Jinan University, Guangzhou 510632, China (e-mail: chenyy@jnu.edu.cn).}
		\thanks{L. Hanzo is with the School of Electronics and Computer Science, University of Southampton, Southampton, SO17 1BJ, U.K. (email: lh@ecs.soton.ac.uk).}
	}

	\maketitle
	\begin{abstract}
		Both the Mobile edge computing (MEC)-based and fog computing (FC)-aided Internet of Vehicles (IoV) constitute promising paradigms of meeting the demands of low-latency pervasive computing. To this end, we construct a dynamic NOMA-based computation offloading scheme for vehicular platoons on highways, where the vehicles can offload their computing tasks to other platoon members. To cope with the rapidly fluctuating channel quality, we divide the timeline into successive time slots according to the channel's coherence time. Robust computing and offloading decisions are made for each time slot after taking the channel estimation errors into account. Considering a certain time slot, we first analytically characterize both the locally computed source data and the offloaded source data as well as the energy consumption of every vehicle in the platoons. We then formulate the problem of minimizing the long-term maximum task queue by optimizing the allocation of both the communication and computing resources. To solve the problem formulated, we design an online algorithm based on the classic Lyapunov optimization method and successive convex approximation (SCA) method. Finally, the numerical simulation results characterize the performance of our algorithm and demonstrate its advantages both over the local computing scheme and the orthogonal multiple access (OMA)-based offloading scheme. 
	\end{abstract}
	
	
	\IEEEpeerreviewmaketitle
	
	\section{Introduction}
	\IEEEPARstart{T}{he} Internet of Vehicles (IoV) has attracted substantial attention both in industry and academia with the objective of improving traffic efficiency and improving road safety \cite{Alam2015,Contreras2018,WZhang2017,Ning2019}, by sharing information among vehicles and roadside units (RSUs) within their coverage area. Typical applications include safety-related, driving-assistance and passenger entertainment services \cite{Feng2017}. The safety-critical tasks are mainly executed locally by the on-board unit for eliminating any queuing and communications delays as well as transmission impairments. By contrast, some applications rely on large amounts of data processing and/or require low latency. The specific services can include vehicular sensing \cite{Wang_J2018}, low-latency lip-synchronized video streaming \cite{Zhou_G2020}, flawless interactive gaming \cite{KZhang2017}, and other Internet-of-vehicles as well as social networking services\cite{LiWang, JieHu}.  In a nutshell, the demands for computing resources have increased quite dramatically owing to the augmented security requirements and expectations of passengers, especially when aiming for taking into account the environmental awareness of all other vehicles in their decision-making.
	
	To overcome these challenges, mobile edge computing (MEC) and fog computing (FC) constitute a pair of promising design paradigms for offloading computing-intensive tasks to the edge of the mobile networks and fog nodes \cite{KZhang2017,Hou2016,You2017,Huang2017}. As a benefit of closer proximity to computing servers, the data transmission delay can be significantly reduced in MEC-aided or FC-assisted vehicular networks, which substantially improves their efficiency.  In urban scenarios, the computing nodes are usually part of the roadside units (RSUs) \cite{Du2019,Sorkhoh2019}. However, it is costly to deploy lots of RSUs along all highways. As a design alternative, we consider vehicular platoons, which exhibit slowly varying formations \cite{Jia2016,Zeng2019}, where vehicles may offload their tasks to their platoon members having fewer or less demanding computing tasks. This also supports the option of formulating decisions based on the joint environmental awareness of the nodes in the platoon, taking into account the sensory information of all vehicles in it.
	
	In vehicular communications, spectrum shortage is another issue when considering data transmission among vehicles requesting computation offloading and those assisting in offloaded processing. In this context, non-orthogonal multiple access (NOMA) can serve several users within the same time and frequency resource block, which further improves the bandwidth efficiency \cite{Ding2017}, as evidenced in \cite{ChenY2017,YLiu2018,Ding2019T}.
	
	Motivated by the above-mentioned aspects, we construct a NOMA-based computation offloading scheme for targeted applications. Specifically, the vehicles in the platoon can offload computing tasks to the neighboring vehicles or retrieve results from other platoon members having less computing tasks by relying on the dedicated short-range communications (DSRC) protocol or on the cellular-vehicle-to-everything (C-V2X) protocol \cite{Naik2019}. Given the potentially high speed of vehicles, we assume an uncorrelated Rayleigh fading channel model and conceive an online algorithm by making offloading decisions by allocating the communication resources according to the near-instantaneous channel conditions and additionally considering the task queues of the vehicles in each time slot. The channel estimation errors are also taken into account in the vehicular platoons.
	
	The main contributions of our paper are summarized as follows: 
	\begin{itemize}
		\item We propose a dynamic NOMA-based computation offloading scheme	for a vehicular platoon. The problem of minimizing the average maximum task queue length at vehicles is formulated for optimizing the offloading decisions and the communication resource allocation.
		
		\item We design the computing and offloading decisions on a time-slot by time-slot basis and take the channel estimation errors into account.
		
		\item In order to solve the problem formulated, we propose an online algorithm based on the classic Lyapunov optimization method \cite{Neely2010} and on the successive convex approximation (SCA) method \cite{Marks1978}.
		
		\item Finally, the advantages of NOMA-based computation offloading over both local computing and orthogonal multiple access (OMA)-based offloading are quantified.
	\end{itemize}
	
	The remainder of this paper is organized as follows. We discuss the related literature in Section II. Then, we introduce our system model and formulate the optimization problem considered in Section III. The algorithm scheduling the computation offloading and resource allocation decisions is designed in Section IV. Finally, our simulation results are given in Section V, and Section VI concludes this paper. 
	
	\section{Related Contributions}
	NOMA-based computation offloading has been shown to enhance the performance of MEC and FC systems \cite{Pan2019,Wang2018,Ding2018,Wu2019,Pei2020,Liu2019,Nouri2019,Wei2018,Wang2020,Qian2020}. The existing studies can be divided into two categories according to whether static or dynamic NOMA-based computation offloading is considered in the problem formulation.
	
	\emph{For static NOMA-based computation offloading}, diverse optimization metrics have been investigated without considering dynamic task arrivals and channel variations. To minimize the energy consumption, the authors of \cite{Pan2019} considered a scenario where both the task offloading and result retrieval relied on NOMA. Moreover, the authors of \cite{Wang2018} studied NOMA-based computation offloading among mobile devices, where beamforming was introduced for reducing the interference among different NOMA pairs. To minimize the time delay, Ding \emph{et al.} \cite{Ding2018} investigated three multiple access schemes -- OMA, pure NOMA, and hybrid NOMA -- for MEC offloading in a two-users scenario. Furthermore, the authors of \cite{Wu2019} extended the NOMA-based computation offloading philosophy to a general multi-users scenario, where both task offloading and result retrieval relied on NOMA. Pei \emph{et al.} \cite{Pei2020} studied secure NOMA-based computation offloading in the presence of eavesdroppers. To maximize the total transmission rate, the authors of \cite{Liu2019} investigated the scenario where a host device offloaded its latency-constrained computation task to the surrounding cooperative devices by combining both non-orthogonal downlink and uplink transmissions. 
	However, these static NOMA-based computation offloading schemes are not suitable for the platoons considered in the face of dynamic task arrival.
	
	\emph{For dynamic NOMA-based computation offloading}, both the channel fluctuations and dynamic task arrivals are considered. The authors of \cite{Nouri2019} and \cite{Wei2018} investigated in the multi-user NOMA uplink computation offloading and multi-server aided downlink NOMA computation offloading, respectively. The classic Lyapunov optimization method was used for converting the long-term utility optimization problem into an online optimization problem in both these studies. As a further advance, the authors of \cite{Wang2020} considered both the long-term and short-term system behavior, where machine learning based prediction and Lyapunov optimization were adopted, respectively. Qian \emph{et al.} \cite{Qian2020} investigated a multi-task multi-access MEC system, where deep reinforcement learning was used for finding near-optimal offloading solutions for time-varying channel realizations. However, these treatises have assumed perfect channel state information (CSI), which is impractical especially in vehicular scenarios, since the movement of wireless terminals will impact the CSI acquisition accuracy.
	
	Specifically, we have summarized the related works and compared our contribution in Table I. By taking both the dynamic channel environment as well as dynamic task arrival and realistic imperfect channel estimation into account, we construct a dynamic cooperative NOMA-based computation offloading scheme for a vehicular platoon. Both the computation offloading and communication resource allocation are optimized for all platoon members according to the near-instantaneous channel state and task queues at the vehicles. NOMA-based computation offloading is considered among all platoon members, and no other edge computing servers are involved in the scenario considered.

	\begin{table*}		
		\centering
		\renewcommand{\arraystretch}{1.3}
		\centering
		\caption{Contrasting our contribution to the literature.}
		\label{Table1}
		\begin{tabular}{l|c|c|c|c|c|c}
			\hline
			  & {[25],[26]}  & [27]--[29] &[30]& [31] & {[32]--[34]} & {\bf{Our work}} \\
			\hline
			\hline
			{NOMA-based offloading} & $\checkmark$ & $\checkmark$  & $\checkmark$ &$\checkmark$ &$\checkmark$ & $\boldmath{\checkmark}$  \\
			\hline
			{Cooperative offloading} &  &   &  $\checkmark$& && $\boldmath{\checkmark}$  \\
			\hline
			{Energy consumption} & $\checkmark$ & & $\checkmark$ &$\checkmark$ &$\checkmark$ & $\boldmath{\checkmark}$  \\
			\hline
			{Time delay}  & & $\checkmark$   &  $\checkmark$ &&$\checkmark$ &  $\boldmath{\checkmark}$  \\
			\hline
			{Dynamic offloading}  &  & & &$\checkmark$ &$\checkmark$  &  $\boldmath{\checkmark}$   \\
			\hline
			{Imperfect channel estimation} &  &   &  & & &  $\boldmath{\checkmark}$ \\
			\hline	
		\end{tabular}
	\end{table*}

	\begin{figure}[!t]
		\centering
		\includegraphics[width=0.45\textwidth]{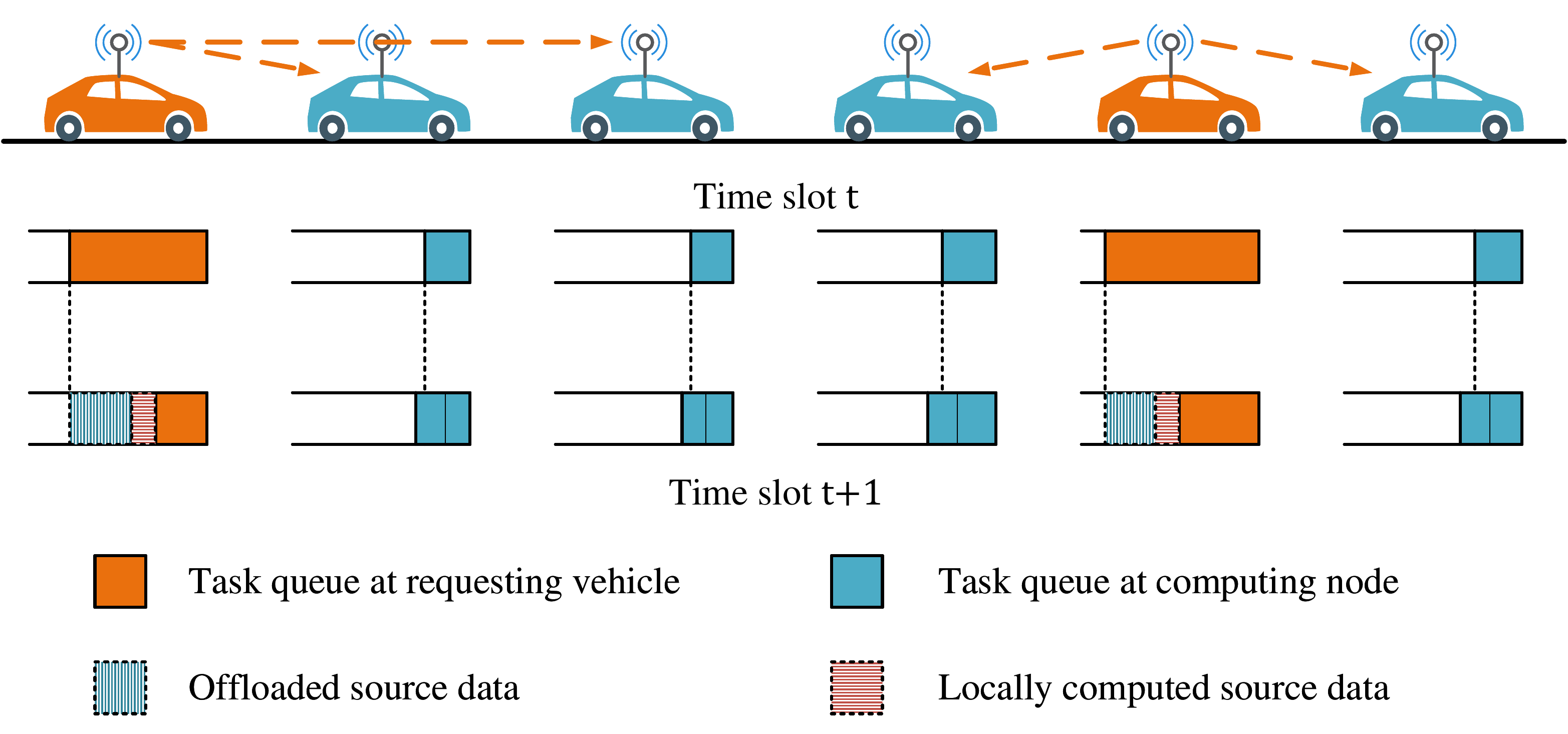}
		\caption{The considered vehicular platoon system.}
		\label{system_structure}
	\end{figure}
	
	\section{System Model And Problem Formulation}
	\subsection{System Overview}
	As shown in Fig. \ref{system_structure}, we consider a steadily moving vehicular platoon on the highway. Denote vehicles within the platoon as $\mathcal{K}=\{1, 2, \cdots, k, \cdots, K\}$ with $K$ representing the number of vehicles, which can partially or fully offload their computation-intensive tasks to other platoon members. On the other hand, they can also serve as computing nodes processing the tasks offloaded by other platoon members to them. To differentiate the two roles of each vehicle, we denote the computing node set as $\mathcal{N}=\{1, 2, \cdots, n, \cdots, N\}$ with $N=K$. The distance  between any two adjacent vehicles in the platoon is assumed to be equal.  We note that this assumption is reasonable because the commercially available cooperative adaptive cruise control (CACC) of \cite{HuangR2017} is already capable of satisfying this. Furthermore, the distance is velocity-dependent \cite{Jia2014, Treiber2000}
	\begin{equation}
		d_v=\frac{d_0+vt_{0}}{\sqrt{1-\left({v}/{v_m}\right)^4}},
	\end{equation}
	where $d_0$ is the minimum intra-platoon spacing, $v$ is the driving velocity, $t_0$ is the desired time headway, and $v_m$ is the maximum speed.
	
	We divide the timeline into successive time slot (TSs) according to the channel's coherence time, which are denoted by
	$\mathcal{T}=\left\{1,2,\cdots,t,\cdots\right\}$. The duration of a TS is denoted by $\tau$. Furthermore, the channel quality at TSs is assumed to remain time-invariant and be uncorrelated with that of other TSs. Each vehicle has some computation-intensive tasks. We assume that the arrival of such tasks follows Poisson distribution with an arrival rate of $\lambda$, and the data size of the tasks follows a uniform distribution
	$\theta_k(t)\sim{\rm{U}}\left(\theta_k^{\textrm{min}},\theta_k^{\textrm{max}}\right)$. Usually, the time interval between two successive computation-intensive tasks tends to be long, which indicates that $\theta_k(t)$ equals to $0$ in most TSs.
	
	In each TS, the specific vehicles having heavy computational burden can offload parts of their tasks to other vehicles by relying on the non-orthogonal downlink and can execute parts of their tasks locally. Meanwhile, some vehicles receive the offloaded tasks and push it into their task buffer. These tasks are then executed in the following TS.
	
	\subsection{NOMA Offloading}
	In this subsection, we will elaborate on the process of data transmission between the requesting vehicle and the computing nodes. Upon taking the channel estimation errors into account in the platoon \cite{Chen2020}, the actual channel coefficient can be written as
	\begin{equation}
		h_{kn}(t)=\hat{h}_{kn}(t)+\Delta h_{kn}(t),
	\end{equation}
	with $\Delta h_{kn} \sim \mathcal{CN}\left(0,\sigma_h^2\right)$, where $\sigma_h^2$ is the variance of the channel estimation error. Under the assumption of uncorrelated Rayleigh fading, the estimated channel coefficient $\hat{h}_{kn}(t)$ and the channel estimation error $\Delta h_{kn}(t)$ remain time-invariant during a TS, but they vary randomly and independently between consecutive TSs. Moreover, the estimated channel coefficient $\hat{h}_{kn}$ can be modeled by \cite{Chen2020}
	\begin{equation}
		\hat{h}_{kn}(t)=\sqrt{G}(d_{kn})^{-\frac{\phi}{2}}g_{kn}(t),
	\end{equation} 
	where $G$ is the path-loss factor, $d_{kn}$ is the distance between vehicle $k$ and computing node $n$, $\phi$ is the path-loss exponent, and $g_{kn}$ reflects the small-scale fading, which obeys a complex Gaussian distribution with zero mean and unit variance. The distance between vehicle $k$ and computing node $n$ can be  further written as $d_{kn}=|k-n|d_v$. 
	
	Let us consider the task offloading of vehicle $k$. Assume that the computing node $n$ $(n \ne k)$ receives the superimposed NOMA signal from the requesting vehicle $k$, which is formulated as
	\begin{align}
		r_{kn}(t) &= \left(\hat{h}_{kn}(t)+\Delta h_{kn}(t)\right)\sum \limits_{i\in \mathcal{N}} \sqrt{p_{ki}(t)}x_{ki}+\varpi_{kn} \nonumber\\
		&=\hat{h}_{kn}(t)\sqrt{p_{kn}(t)}x_{kn}+\Delta h_{kn}(t)\sum \limits_{i\in\mathcal{N}}  \sqrt{p_{ki}(t)}x_{ki} \nonumber\\
		&\quad+\hat{h}_{kn}(t)\sum \limits_{i\in \mathcal{N}\setminus\{n\}}\sqrt{p_{ki}(t)}x_{ki}+\varpi_{kn} (t),
	\end{align}
	where $p_{ki}(t)$ denotes the power allocated to transmit source data from vehicle $k$ to computing node $i$ in TS $t$,  $x_{ki}$ with $\mathbb{E} \left[{|x_{ki}|^2}\right]=1$ denotes the signal transmitted from vehicle $k$ to computing node $i$ and $\varpi_{kn}(t)$ is the complex additive white Gaussian noise process having a power spectral density $N_0$. It is noted that $p_{ki}(t)=0$ indicates that the computing task at vehicle $k$ is not offloaded to computing node $i$ in TS $t$, and $p_{kk}(t)=0$ holds for all vehicles.
	
	The successive interference cancellation decoding process \cite{Liu_Y2017} is adopted to retrieve the transmitted data at the computing nodes. To be specific, the computing node first decodes the signal transmitted to vehicles having better channel quality, and then remodulates the decoded signal followed by subtracting it from the received composite signal.  Therefore, the signal to interference plus noise ratio (SINR) at computing nodes is formulated as
	\begin{equation}\label{SINR-offloading}
		{\Gamma _{kn}(t)} =  \frac{p_{kn}(t)|\hat{h}_{kn}(t)|^2}{H_n\left[\mathbf{p}_k(t),w_k(t)\right]+ |\Delta h_{kn}(t)|^2\sum\limits_{i\in\mathcal{N}}p_{ki}(t)},
	\end{equation} 
	where the function $H_n\left[\mathbf{p}_k(t),w_k(t)\right]$ is defined as 
	\begin{equation*}
		\begin{split}
			&H_n\left[\mathbf{p}_k(t),w_k(t)\right]\\
			&=|\hat{h}_{kn}(t)|^2\sum \limits_{i\in \mathcal{N}}p_{ki}(t)\mathbb{I}{\left[|\hat{h}_{ki}(t)|^2>|\hat{h}_{kn}(t)|^2\right]}+w_k(t)N_0
		\end{split}
	\end{equation*}
	with the indicator function $\mathbb{I}(\cdot)$ given by
	\begin{equation}
		\mathbb{I}{\left[|\hat{h}_{ki}(t)|^2>|\hat{h}_{kn}(t)|^2\right]}=\begin{cases}
			1,\ \textrm{if} \ |\hat{h}_{ki}(t)|^2>|\hat{h}_{kn}(t)|^2, \\
			0,\ \textrm{otherwise}.
		\end{cases}
	\end{equation}

	Moreover, the frequency band occupied by different vehicles for offloading their task is assumed to be orthogonal, which indicates that the constraint of $\sum \nolimits_{k \in \mathcal{K}}w_k(t)\le W_0$ with $W_0$ denoting the total available bandwidth holds. Then the achievable data rate of offloading data from vehicle $k$ to computing node $n$ in TS $t$ is 
	\begin{equation}\label{Capacity-offloading}
		R_{kn}^{\textrm{achi}}(t) = w_k(t)\cdot{\log _2}\left[ {1 + \Gamma_{kn}(t)} \right].
	\end{equation} 
	
	We denote the transmission data rate of vehicle $k$ to computing node $n$ in TS $t$ by  $R_{kn}(t)$. Then, the data offloading $D_k^{\textrm{off}}(t)$ in TS $t$ can be characterized by
		\begin{equation}\label{offloaded_data}
			D_k^{\textrm{off}}(t)=\sum_{n\in \mathcal{N}\setminus\{k\}} \pi_{kn}(t)R_{kn}(t)\tau,
		\end{equation}
	where $\pi_{kn}(t)$ is a binary variable with $\pi_{kn}(t)=1$ representing successful data offloading from vehicle $k$ to vehicle $n$, and $\pi_{kn}(t)=0$, otherwise. Moreover, we have the following equations
	\begin{equation}
		\begin{cases}
			\Pr\left\{\pi_{kn}(t)=0\right\}=\eta_0,\\
			\Pr\left\{\pi_{kn}(t)=1\right\}=1-\eta_0,
		\end{cases}
	\end{equation}
	where $\Pr(\cdot)$ denotes the probability of an event, and $\eta_0$ represents the outage probability, which is formulated as
	\begin{align}\label{outage_probability}
		\eta_0&=\textrm{Pr}\left\{R_{kn}(t)>R_{kn}^{\textrm{achi}}(t)\right\}\nonumber \\
		&=\textrm{Pr}\left\{R_{kn}(t) > w_k(t)\log_2 {\left(1+\Gamma_{kn}(t)\right)}\right\}\nonumber \\
		&=\textrm{Pr}\left\{|\Delta h_{kn}(t)|^2> \frac{\dfrac{p_{kn}(t)|\hat{h}_{kn}(t)|^2}{2^{{R_{kn}(t)}/{w_k(t)}}-1}-H_n\left[\mathbf{p}_k(t),w_k(t)\right]}{\sum\limits_{i\in\mathcal{N}}p_{ki}(t)}\right\} \nonumber \\
		&=\exp\left\{-\frac{1}{\sigma_h^2}\frac{\dfrac{p_{kn}(t)|\hat{h}_{kn}(t)|^2}{2^{{R_{kn}(t)}/{w_k(t)}}-1}-H_n\left[\mathbf{p}_k(t),w_k(t)\right]}{\sum\limits_{i\in\mathcal{N}}p_{ki}(t)}\right\}.
	\end{align}

	Alternatively, we can express the transmission rate $R_{kn}(t)$ in terms of the outage probability $\eta_0$ as
	\begin{align}\label{rate}
		&R_{kn}(t)= \nonumber \\
		&w_{k}(t) \log_2\left\{1+\dfrac{p_{kn}(t)|\hat{h}_{kn}(t)|^2}{H_n\left[\mathbf{p}_k(t),w_k(t)\right]-\ln(\eta_0)\sigma_h^2\sum\limits_{i\in\mathcal{N}}p_{ki}(t)}\right\}.
	\end{align}

	Considering the relatively small size of computation results, we ignore the data rate requirement of retrieving the results from the corresponding computing nodes. Then, the energy consumption during this offloading process can be formulated as  
	\begin{equation}
		E_k^{\textrm{off}}(t)= \sum\limits_{n\in\mathcal{N}}{p_{{kn}}(t)}\tau.
	\end{equation}
	
	\subsection{Local Computation Model}
	Here, we briefly introduce the computational model of \cite{Mao2017} used for characterizing the computing capability of vehicles. The maximum operating frequency (in CPU cycles per second) of all vehicles is assumed to be equal and it is denoted by $f_m$. Leveraging dynamic voltage and frequency scaling techniques, the operating frequency can be dynamically adjusted according to the current computing burden. We denote the clock-frequency (in CPU cycles per second) at vehicle $k$ in TS $t$ as $f_k(t)$. Then the locally computed data during TS $t$ can be written as
	\begin{equation}\label{local_exectution}
		D_k^{\textrm{comp}}(t)=\frac{f_k(t)}{\epsilon_k}\tau,
	\end{equation}
	where $\epsilon_k$ is the computational workload reflecting the number of CPU cycles required for processing a single input bit.
	
	Furthermore, according to the well-established computational model of \cite{You2017}, the energy consumption of the local computing at vehicle $k$ in TS $t$ is formulated as
	\begin{equation}
		E_k^{\textrm{comp}}(t)=\xi_kf_k^3(t)\tau,
	\end{equation}
	where $\xi_k$ is the energy efficiency coefficient determined by the architecture of the processor used at vehicle $k$. 
	
	Therefore, the total energy consumption of vehicle $k$ in TS $t$ is formulated as 
	\begin{equation}\label{energy_total}
		E_k^{\textrm{total}}(t)={\sum\limits_{n\in\mathcal{N}} {{p _{kn}(t)}} } \tau+\xi_kf_k^3(t)\tau.
	\end{equation}
	
	\subsection{Problem Formulation}
	Considering an individual vehicle $k$, the data that departed from the task queue includes the offloading data $D_k^{\textrm{off}}(t)$ and the locally executed data $D_k^{\textrm{comp}}(t)$ in TS $t$. Meanwhile, for the computing node $k$, the data that arrived at the input of the task queue is comprised of the data  received from other vehicles 
	\begin{equation}\label{received_data}
		A_k^{\textrm{rece}}(t)=\sum_{i\in \mathcal{K}\setminus\{k\}} \pi_{ik}(t)R_{ik}(t)\tau,
	\end{equation}
	and the newly generated task $\theta_k(t)$ in TS $t$.
	
	Assume that each vehicle is equipped with sufficient storage to store the source data to be processed and the length of the task queue at vehicle $k$ in TS $t$ is denoted by $Q_k(t)$. Then the associated computing task queue is formulated as
	\begin{align}\label{task_queue}
		Q_k(t+1)&=\left[Q_k(t)-D_k^{\textrm{off}}(t)-D_k^{\textrm{comp}}(t)\right]^{+}+A_k^{\textrm{rece}}(t)+\theta_k(t)\nonumber\\
		&\triangleq Q_k^{\mathrm{left}}(t)+\theta_k(t),
	\end{align}
	where $\left[x\right]^{+}$ equals the larger one between $0$ and $x$, and $Q_k^{\mathrm{left}}(t)$ denotes the outstanding task queue not including the newly generated task at the end of TS $t$.
	
	Given that we aim for lowest possible delay, the problem to be solved is to minimize the average maximum outstanding task queue length $Q_k^{\mathrm{left}}(t)$,
	which can be written as 
		\begin{alignat}{3} 
			&\mathcal{P}1: &\ & \mathop {\min}\limits_{\mathbf{P}(t),\mathbf{w}(t),\mathbf{f}(t)} \lim\limits_{T \to \infty} \frac{1}{T}\sum_{t=1}^{T}\max\limits_{k \in \mathcal{K}}\left\{{\mathbb{E}\left[Q_k^{\mathrm{left}}(t)\right]}\right\}\label{op1}\\
			&\textrm {s.t.}& &f_k(t)\le f_m,\quad \forall k \in \mathcal{K},\  \forall t \in \mathcal{T}, \tag{\ref{op1}a} \label{op1_c1}\\
			& & &\sum\limits_{n \in \mathcal{N}}p_{kn}(t) \le P_0, \quad \forall k \in \mathcal{K}, \ \forall t \in \mathcal{T}, \tag{\ref{op1}b} \label{op1_c2}\\
			& & &\sum\limits_{k\in\mathcal{K}} {w_k(t)}  \le W_0,  \quad  \forall t \in \mathcal{T}, \tag{\ref{op1}c} \label{op1_c3}\\
			& & &\lim\limits_{T \to \infty}\sum_{t=1}^{T}\frac{\mathbb{E}\left[E_k^{\textrm{total}}(t)\right]}{T}\le E_0, \quad \forall k \in \mathcal{K}, \tag{\ref{op1}d} \label{op1_c4}
		\end{alignat}
where the optimization parameters $\mathbf{P}(t)$, $\mathbf{w}(t)$ and $\mathbf{f}(t)$ are represented by a matrix or vector holding the transmit power, the frequency band and the CPU clock-frequency of all vehicles in TS $t$, which are given by
	\begin{equation*}
		\arraycolsep=1.2pt \def\arraystretch{1.6}
		\left\{\begin{array}{lll}
			\mathbf{P}(t)&=&\left[\mathbf{p}_1(t),\ \mathbf{p}_2(t),\ \cdots,\ \mathbf{p}_k(t),\ \cdots,\ \mathbf{p}_K(t)\right]^T,\\
			\mathbf{w}(t)&=&\left[{w}_1(t),\  {w}_2(t),\ \cdots,\ {w}_k(t),\ \cdots,\ {w}_K(t)\right],\\
			\mathbf{f}(t)&=&\left[{f}_1(t),\ {f}_2(t),\ \cdots,\ {f}_k(t),\ \cdots,\ {f}_K(t)\right].
		\end{array}\right.
	\end{equation*}
	The vector $\mathbf{p}_k(t)$ can be further written as $\mathbf{p}_k(t)=\left[p_{k1}(t),\ p_{k2}(t), \ \cdots, \ p_{kn}(t), \  \cdots, \ p_{kN}(t)\right]^T$. 
	
	Constraints (\ref{op1_c1}), (\ref{op1_c2}) and (\ref{op1_c4}) reflect that the processing frequency, transmit power and average energy consumption of each vehicle must not exceed their corresponding maximum value $f_m$, $P_0$, and $E_0$, respectively. Constraint
	(\ref{op1_c3}) guarantees that the sum of occupied bandwidth for all vehicles is less than or equal to the total available frequency bandwidth.
	
	\section{Algorithm Design}
	The problem $\mathcal{P}1$ of (\ref{op1}) satisfies the long-term average energy consumption, but reliable decisions have to be made also for each TS. The objective function (OF) and its parameters consider different timescales. Therefore, the long-term OF and long-term constraints have to be projected into each TS for tackling the problem in (\ref{op1}). In this context, the Lyapunov optimization method of \cite{Neely2010} constitutes a powerful tool of constructing a new OF, while ensuring that all long-term objectives are reflected. As a benefit, the transformed problem can be solved by only considering the current network state in each TS. Then, the classic SCA method has been adopted for solving the non-convex transformed problem. We elaborate the algorithm design as follows.

	\subsection{Lyapunov Optimization}
		We now proceed by first reformulating the original optimization task of (\ref{op1}) into a new online optimization problem based on the Lyapunov optimization method, where the system decisions $\mathbf{P}(t),\ \mathbf{w}(t)$, and $\mathbf{f}(t)$ are determined solely by the system state of the current TS. The general procedures of converting this kind of dynamic optimization problem into an online optimization problem by leveraging Lyapunov's method are summarized as follows:
	\begin{itemize}
		\item \textbf{Step 1}: Define the Lyapunov function that involves all actual or virtual queues constructed to reflect the current state of them, such as the average energy consumption of all vehicles in our model considered.
		\item \textbf{Step 2}: The Lyapunov drift is appropriately adjusted for guaranteeing the stability of the queues constructed in Step 1. For example, the Lyapunov drift will be increased when the energy consumption of vehicles becomes large.
		\item \textbf{Step 3}: The Lyapunov drift-plus-penalty function regarded as the new optimization objective is formulated by appropriately weighting the conditional Lyapunov drift and the primary optimization OF.
	\end{itemize}
	
	Based on the above steps, the long-term utility and long-term constraints can be removed from the problem formulated, which simplifies the algorithmic design.  Furthermore, the performance gap between the optimal solution of the primary problem and the solution of the reformulated problem can be rendered arbitrarily small, as proven in \cite{Neely2010}. Here, we adopt the Lyapunov-based method for designing an online algorithm to approximately apportion the computing and communication resources for all vehicles in the vehicular platoon.
	
	Firstly, similar to \cite[\emph{Lemma} 1]{Wei2018}, the constraint (\ref{op1_c4}) can be replaced by 
	\begin{equation}
		\lim\limits_{T \to \infty}\frac{\mathbb{E}\left[U_k(t)\right]}{T} = 0,\quad \forall k \in \mathcal{K},
	\end{equation}
	where $U_k(t)$ is a virtual energy consumption queue defined as 
	\begin{equation}\label{energy_queue}
		U_k(t+1)=\left[U_k(t)+E_k^{\mathrm{total}}(t)-E_0\right]^{+}.
	\end{equation}
	Denote all virtual energy consumption queues as $\mathbf{\Theta}(t)=\left[U_1(t),\ U_2(t),\ \cdots,\ U_K(t)\right]$. Then the Lyapunov function is defined by 
	\begin{equation}
		L\left[\mathbf{\Theta}(t)\right]=\frac{1}{2}\sum_{k \in \mathcal{K}}U_k^2(t),
	\end{equation}
	while the Lyapunov drift can be expressed as
	\begin{equation}
		\Delta\left[\mathbf{\Theta} (t)\right]=L\left[\mathbf{\Theta}(t+1)\right]-L\left[\mathbf{\Theta}(t)\right].
	\end{equation}
	Then, the Lyapunov drift-plus-penalty function is formulated as
	\begin{equation}
		\Delta_V\left[\mathbf{\Theta} (t)\right]=\mathbb{E}\left\{\Delta\left[\mathbf{\Theta} (t)\right]+V\max\limits_{k \in \mathcal{K}}\left\{{\mathbb{E}\left[Q_k^{\mathrm{left}}(t)\right]}\right\}|\mathbf{\Theta}(t)\right\},
	\end{equation}
	where $V$ is a non-negative weight reflecting the tradeoff between the system's delay vs. energy consumption and its queue stability. A larger value of $V$ indicates that reducing the outstanding task queue $Q_k^{\mathrm{left}}(t)$ is more important than the energy consumption of the system. Therefore, the choice of the weight $V$ relies on the specific requirements of the vehicular platoons considered.
	
	\begin{lemma}
		For arbitrary feasible $\mathbf{P}(t),\ \mathbf{w}(t),\ \mathbf{f}(t)$, the Lyapunov drift-plus-penalty function is upper bounded by
		\begin{align}\label{V_Upper}
			\Delta_V(\mathbf{\Theta} (t))&\le \Phi+\mathbb{E}\left[\sum_{k \in \mathcal{K}} U_k(t)\left[E_k^{\mathrm{total}}(t)-E_0\right]|\mathbf{\Theta}(t)\right] \nonumber\\
			&\quad+\mathbb{E}\left[V\max\limits_{k \in \mathcal{K}}\left\{{Q_k^{\mathrm{left}}(t)}\right\}|\mathbf{\Theta}(t)\right],
		\end{align}
	where $\Phi$ is a constant.
	\end{lemma} 
	
	The proof is relegated to Appendix A for retaining the flow of our arguments. 
	
	Then, the primary optimization problem $\mathcal{P}1$ of (\ref{op1}) can be transformed into minimizing the upper bound of the Lyapunov drift-plus-penalty function, which is formulated as 
	\begin{alignat}{3} 
		&\mathcal{P}2: &\ &  \min_{\begin{subarray}{c}
		\mathbf{P}(t),\mathbf{w}(t)\\ \mathbf{f}(t),Z(t)
		\end{subarray}} \sum_{k \in \mathcal{K}} U_k(t)\left[E_k^{\mathrm{total}}(t)-E_0\right]+VZ(t)\label{op2}\\
		&\textrm {s.t.}& &Z(t)\ge{\mathbb{E}\left[Q_k^{\mathrm{left}}(t)\right]},\quad \forall k \in \mathcal{K},\tag{\ref{op2}a}\\
		& & &(\ref{op1_c1}),(\ref{op1_c2})\ \textrm{and} \ (\ref{op1_c3}) \ ,\notag
	\end{alignat}
	where we introduce the optimization parameter $Z(t)$ to deal with the maximization that appeared in the OF.
	
	Upon combining Eqs. (\ref{offloaded_data}), (\ref{local_exectution}), (\ref{received_data}) and (\ref{task_queue}), we obtain 
	\begin{align}\label{Queue_min}
		&\mathbb{E}\left\{Q_k^{\mathrm{left}}(t)\right\}=\mathbb{E}\left\{\left[Q_k(t)-D_k^{\textrm{off}}(t)-D_k^{\textrm{comp}}(t)\right]^{+}+A_k^{\textrm{rece}}(t)\right\} \nonumber\\
		&\ge Q_k(t)+\mathbb{E}\left\{A_k^{\textrm{rece}}(t)-D_k^{\textrm{off}}(t)-D_k^{\textrm{comp}}(t)\right\} \nonumber \\
		&=Q_k(t)+\sum_{i\in \mathcal{K}\setminus\{k\}} \left[1-\eta_0\right]R_{ik}(t)\tau \nonumber \\
		&\quad-\sum_{n\in \mathcal{N}\setminus\{k\}} \left[1-\eta_0\right]R_{kn}(t)\tau-\frac{f_k(t)}{\epsilon_k}\tau\triangleq Q_k^{\mathrm{m}}(t).
	\end{align}
	
	Here, we define a function given by
	\begin{equation}\label{defined_function}
		g_k\left[\mathbf{P}(t),\mathbf{w}(t),\mathbf{f}(t),Z(t)\right] = Q_k^{\mathrm{m}}(t)-Z(t),
	\end{equation}
	which is neither convex nor concave. Then, the constraint (\ref{op2}a) can be replaced by
	\begin{equation}
		g_k\left[\mathbf{P}(t),\mathbf{w}(t),\mathbf{f}(t),Z(t)\right]\le0,\quad \forall k \in \mathcal{K}.
	\end{equation}
	
	\subsection{Successive Convex Approximation (SCA)}
	On the one hand, upon substituting (\ref{energy_total}) into the OF of problem $\mathcal{P}2$, the detailed OF expression can be written as:
	\begin{align}\label{Objective_function}
		&O\left[\mathbf{P}(t),\mathbf{w}(t),\mathbf{f}(t),Z(t)\right]= \nonumber\\
		&=\sum_{k \in \mathcal{K}}U_k(t)\left[{\sum\limits_{n\in\mathcal{N}} {{p _{kn}(t)}} } \tau+\xi_kf_k^3(t)\tau-E_0\right]+VZ(t),
	\end{align}
	which is a convex function.

	On the other hand, upon substituting (\ref{rate}) and (\ref{Queue_min}) into (\ref{defined_function}), the detailed expression is given in (\ref{constrain_max}), which is shown at the top of the next page. Here, we represent the function $g_k\left[\mathbf{P}(t),\mathbf{w}(t),\mathbf{f}(t),Z(t)\right]$ in a more concise way as follows,
	\begin{equation}
		g_k(\mathbf{x})={G_{+}(\mathbf{x})}+{G_{-}(\mathbf{x})},
	\end{equation}
	where $\mathbf{x}=\left[\mathbf{p}_1^T(t),\mathbf{p}_2^T(t),\cdots,\mathbf{p}_K^T(t),\mathbf{w}(t),\mathbf{f}(t),Z(t)\right]$. Moreover, it can be shown that $G_{+}(\mathbf{x})$ is a convex function and ${G_{-}(\mathbf{x})}$ is a concave function. In the following, we adopt the SCA approach \cite{Marks1978} to solve the problem. The main operation of the SCA approach is to successively find a convex restrictive approximation of the non-convex constraint.
	
	\begin{figure}[!t]
		\centering
		\includegraphics[width=0.45\textwidth]{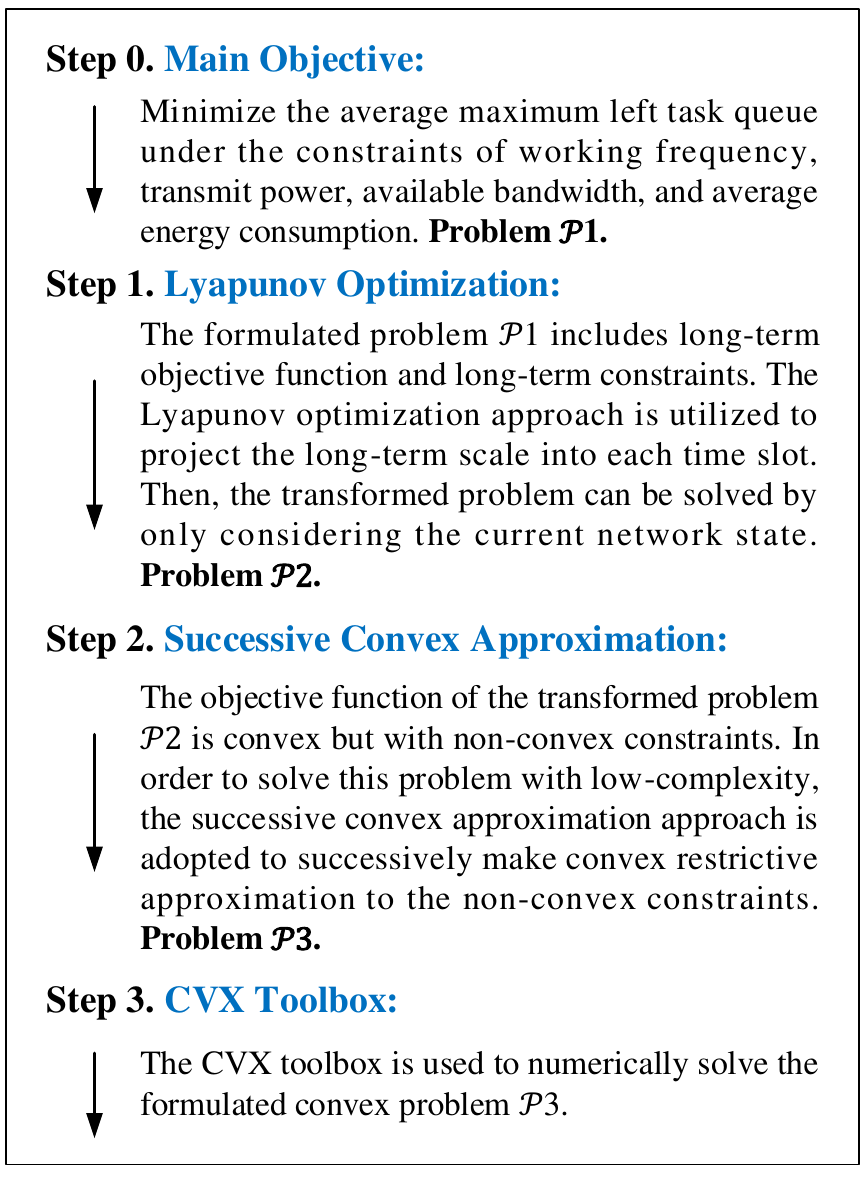}
		\caption{The flow of algorithm design.}
		\label{algorithm_flow}
	\end{figure}

	\begin{figure*}
	\begin{align}\label{constrain_max}
		&g_k\left[\mathbf{P}(t),\mathbf{w}(t),\mathbf{f}(t),Z(t)\right] 
		=Q_k(t)+\sum_{i\in \mathcal{K}\setminus\{k\}} (1-\eta_0)w_{i}(t) \log_2\left\{1+\dfrac{p_{ik}(t)|\hat{h}_{ik}(t)|^2}{H_k\left[\mathbf{p}_i(t),w_i(t)\right]+\ln(1/\eta_0)\sigma_h^2\sum\limits_{n\in \mathcal{N}}p_{in}(t)}\right\}\tau \nonumber\\
		&\quad -\sum_{n\in \mathcal{N}\setminus\{k\}} (1-\eta_0)w_{k}(t) \log_2\left\{1+\dfrac{p_{kn}(t)|\hat{h}_{kn}(t)|^2}{H_n\left[\mathbf{p}_k(t),w_k(t)\right]+\ln(1/\eta_0)\sigma_h^2\sum\limits_{i\in\mathcal{K}}p_{ki}(t)}\right\}\tau-\frac{f_k(t)}{\epsilon_k}\tau-Z(t)\nonumber\\
		&=Q_k(t)-\frac{f_k(t)}{\epsilon_k}\tau-Z(t)-\sum_{i\in \mathcal{K}\setminus\{k\}} (1-\eta_0)w_{i}(t) \log_2\left\{\dfrac{H_k\left[\mathbf{p}_i(t),w_i(t)\right]+\ln(1/\eta_0)\sigma_h^2\sum\limits_{n\in \mathcal{N}}p_{in}(t)}{w_i(t)N_0}\right\}\tau \nonumber \\
		&\underbrace{-\sum_{n\in \mathcal{N}\setminus\{k\}} (1-\eta_0)w_{k}(t) \log_2\left\{\dfrac{p_{kn}(t)|\hat{h}_{kn}(t)|^2+H_n\left[\mathbf{p}_k(t),w_k(t)\right]+\ln(1/\eta_0)\sigma_h^2\sum\limits_{i\in\mathcal{K}}p_{ki}(t)}{w_k(t)N_0}\right\}\tau} _{G_{+}\left[\mathbf{P}(t),\mathbf{w}(t),\mathbf{f}(t),Z(t)\right]} \nonumber\\
		&+\sum_{i\in \mathcal{K}\setminus\{k\}} (1-\eta_0)w_{i}(t) \log_2\left\{\dfrac{p_{ik}(t)|\hat{h}_{ik}(t)|^2+H_k\left[\mathbf{p}_i(t),w_i(t)\right]+\ln(1/\eta_0)\sigma_h^2\sum\limits_{n\in \mathcal{N}}p_{in}(t)}{w_i(t)N_0}\right\}\tau\nonumber \\
		&\underbrace{+\sum_{n\in \mathcal{N}\setminus\{k\}} (1-\eta_0)w_{k}(t) \log_2\left\{\dfrac{H_n\left[\mathbf{p}_k(t),w_k(t)\right]+\ln(1/\eta_0)\sigma_h^2\sum\limits_{i\in\mathcal{K}}p_{ki}(t)}{w_k(t)N_0}\right\}\tau}_{G_{-}\left[\mathbf{P}(t),\mathbf{w}(t),\mathbf{f}(t),Z(t)\right]}.
	\end{align}
	\hrulefill
	\end{figure*}

	Specifically, the updated function $\hat{g}_k(\mathbf{x})$ can be obtained by
	\begin{equation}\label{updating_constraint}
		\hat{g}_k(\mathbf{x})={G_{+}(\mathbf{x})}+\nabla {G_{-}(\mathbf{x}^\star)} \left(\mathbf{x}-\mathbf{x}^\star\right)^T,
	\end{equation}
	where $\mathbf{x}^\star$ is the optimal solution with the last updated constraint $\hat{g}_k(\mathbf{x})\le 0$. Given this approximation, the optimization problem $\mathcal{P}2$ converts into 
		\begin{alignat}{3} 
		&\mathcal{P}3: &\ &  \min_{\begin{subarray}{c}
				\mathbf{P}(t),\mathbf{w}(t)\\ \mathbf{f}(t),Z(t)
		\end{subarray}}  O\left[\mathbf{P}(t),\mathbf{w}(t),\mathbf{f}(t),Z(t)\right]\label{op3}\\
		&\textrm {s.t.}& &\hat{g}_k(\mathbf{x})\le0,\quad \forall k \in \mathcal{K},\tag{\ref{op3}a}\\
		& & &(\ref{op1_c1}),(\ref{op1_c2})\ \textrm{and} \ (\ref{op1_c3}) \ ,\notag
	\end{alignat}
	which is a convex optimization problem. However, it still remains a challenge to obtain the analytical solutions of the problem stated in $\mathcal{P}3$. Here, we adopt the off-the-shelf convex optimization toolbox CVX \cite{Grant2020} to solve this convex problem. Figure \ref{algorithm_flow} illustrates the design flow of the proposed algorithm.
	
	\subsection{Complexity Analysis}
	The SCA algorithm conceived for solving the problem $\mathcal{P}3$ is summarized in Algorithm \ref{algorithm}. Here, we analyze the complexity of the proposed algorithm. According to \cite{Efrem2020}, the complexity of the SCA approach is given by $\mathcal{O}\left((\vartheta/\delta_0)u(K^2+2K+1,3K+1)\right)$, where $\vartheta=O(\mathbf{x}_0^\star)/O(\mathbf{x}^\star)$, $\delta_0$ is the termination threshold as shown in Algorithm \ref{algorithm}, and $u(K^2+2K+1,3K+1)$ denotes the complexity of the problem $\mathcal{P}3$ with $K^2+2K+1$ and $3K+1$ representing the number of variables and constraints, respectively. We note that the optimization problem $\mathcal{P}3$ is a convex one, which can be solved at polynomial complexity. Therefore, the function $u(K^2+2K+1,3K+1)$ is a polynomially increasing function with respect to $K^2+2K+1$ and $3K+1$.
	
	\renewcommand{\algorithmicrequire}{\textbf{Input:}}
	\renewcommand{\algorithmicensure}{\textbf{Output:}}
	\begin{algorithm}[!t]
		\caption{SCA algorithm for problem $\mathcal{P}2$}
		\label{algorithm}
		\begin{algorithmic}[1]
			\REQUIRE ~~\\
			Initial value $\mathbf{x}_0^\star=\left[{\mathbf{p}_{10}^\star(t)}^T,\cdots,{\mathbf{p}_{K0}^\star(t)}^T,\mathbf{w}_0^\star(t),\mathbf{f}_0^\star(t),{Z}_0^\star(t)\right]$;
			\ENSURE ~~\\
			Optimal power allocation ${\mathbf{P}^\star(t)}$;\\
			Optimal spectrum resource allocation $\mathbf{w}^\star(t)$;\\
			Optimal CPU frequency $\mathbf{f}^\star(t)$;
			\STATE Set $i=0$;
			\REPEAT
			\STATE Set $\mathbf{x}^\star=\mathbf{x}_i^\star$;
			\STATE Updating the constraint $\hat{g}_k({\mathbf{x}})$ through equation (\ref{updating_constraint});
			\STATE Updating $\mathbf{x}^\star$ by solving optimization problem $\mathcal{P}3$;
			\STATE Set $i =i+1$ and $\mathbf{x}_i^\star=\mathbf{x}^\star$;
			\UNTIL{$\left|O{\left(\mathbf{x}_{i}^\star\right)}-O{\left(\mathbf{x}_{i-1}^\star\right)}\right|\le\delta_0O{\left(\mathbf{x}_{i-1}^\star\right)}$.}
			\STATE  Output the obtained solution ${\mathbf{P}^\star(t)}$, $\mathbf{w}^\star(t)$, and $\mathbf{f}^\star(t)$.
		\end{algorithmic}
	\end{algorithm}

	\section{ Numerical Results} 
	In this section, we evaluate the performance of the proposed algorithm in the dynamic NOMA-based computation offloading aided platooning system. Unless otherwise stated, the default simulation parameters are set as shown in TABLE \ref{table_simulation}. The average maximum queue length in TS $t$ is defined as $1/t\sum \nolimits_{i=1}^{t}\max\limits_{k \in \mathcal{K}}\left\{Q_k(i)\right\}$. For comparison, we choose the OMA-based offloading system and the local computing system as the benchmarks. In the OMA-based counterpart, the bandwidth allocated to each communication mode pair is orthogonal. Furthermore, to ensure the fairness of comparisons, we have appropriately modified the algorithm designed for NOMA-based computation offloading to make it suitable for its OMA-based counterpart. All simulation results are obtained by	averaging over 10000 independent channel realizations. The number of vehicles $K$ is set to 8.
	
	\begin{table}[!t]
		\renewcommand{\arraystretch}{1.0} 
		\caption{Simulation Parameters}
		\label{table_simulation} 
		\centering 
		\begin{tabular}{l|l} 
			\hline 
			\hline
			\bfseries Parameter & \bfseries Value\\ 
			\hline
			Minimum intra-platoon spacing $d_0$ & $3.0$ m\\
			Time headway $t_0$ & $1.5$ s\\
			Maximum driving speed $v_m$ & $120$ km/h\\
			Computation workload $\epsilon_{k}$  & $40$\\
			Energy coefficient $\xi_k$ & $1\times10^{-27}$\\
			Time interval $\tau$ & $1$ ms\\
			Maximum operating frequency $f_m$ & $1$ GHz\\
			Constant path-loss factor $G$ & $-31.5$ dB\\
			Path-loss exponent $\phi$ & $2$\\
			Noise power spectral density $N_0$ & $-174 \ \textrm{dBm/Hz}$\\
			Maximum transmit power of vehicles $P_0$ & $35$ dBm\\
			Bandwidth $W_0$ & $20$ MHz\\
			Arrival rate $\lambda$ & $1/120$\\
			Lower bound of generated data $\theta_k^{\textrm{min}}$ & $5\times10^5$ bits\\
			Upper bound of generated data $\theta_k^{\textrm{max}}$ & $6\times10^5$ bits\\
			Outage probability $\eta_0$ & $0.1$\\
			Termination threshold $\delta_0$ & $1\times10^{-6}$\\
			\hline
			\hline 
		\end{tabular} 
	\end{table} 

\begin{figure}[!t]
		\centering
		\includegraphics[width=0.45\textwidth]{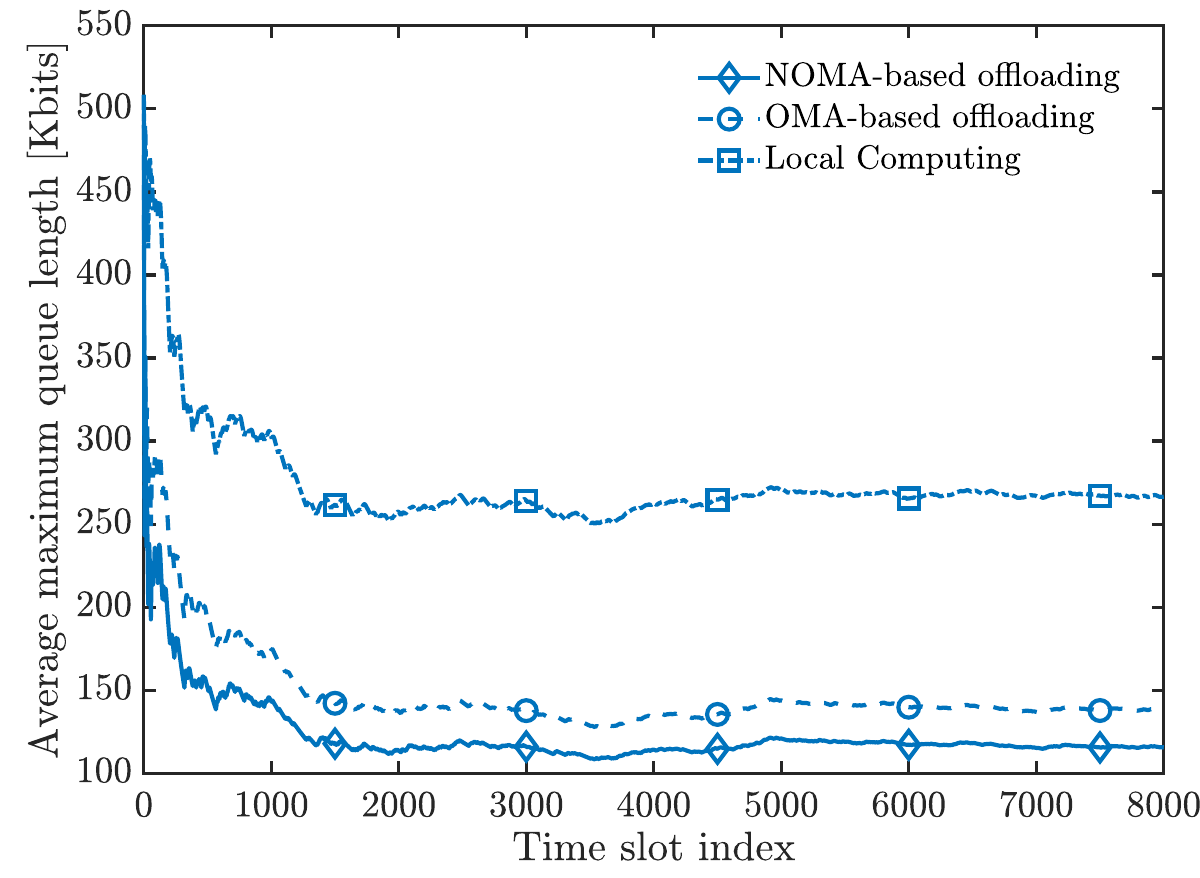}
		\caption{Average maximum queue length versus the TS index under different offloading strategies evaluated by simulations.}
		\label{Queue_variation}
	\end{figure}

	Figure \ref{Queue_variation} shows the average maximum queue length in different TSs under different offloading schemes. The velocity $v$, the channel estimation error variance $\sigma_h^2$, and the weighting factor $V$ are set to $60$ km/h, $1\times10^{-16}$, and $1\times10^{-8}$, respectively. Over time, the average queue length converges to a specific value. It can be inferred from Fig. \ref{Queue_variation} that cooperative computation offloading among platoon members through NOMA or OMA reduces the average computing burden of the individual vehicles. Furthermore, the NOMA-based computation offloading scheme  outperforms its OMA-based counterpart.
	
\begin{figure}[!t]
		\centering
		\includegraphics[width=0.45\textwidth]{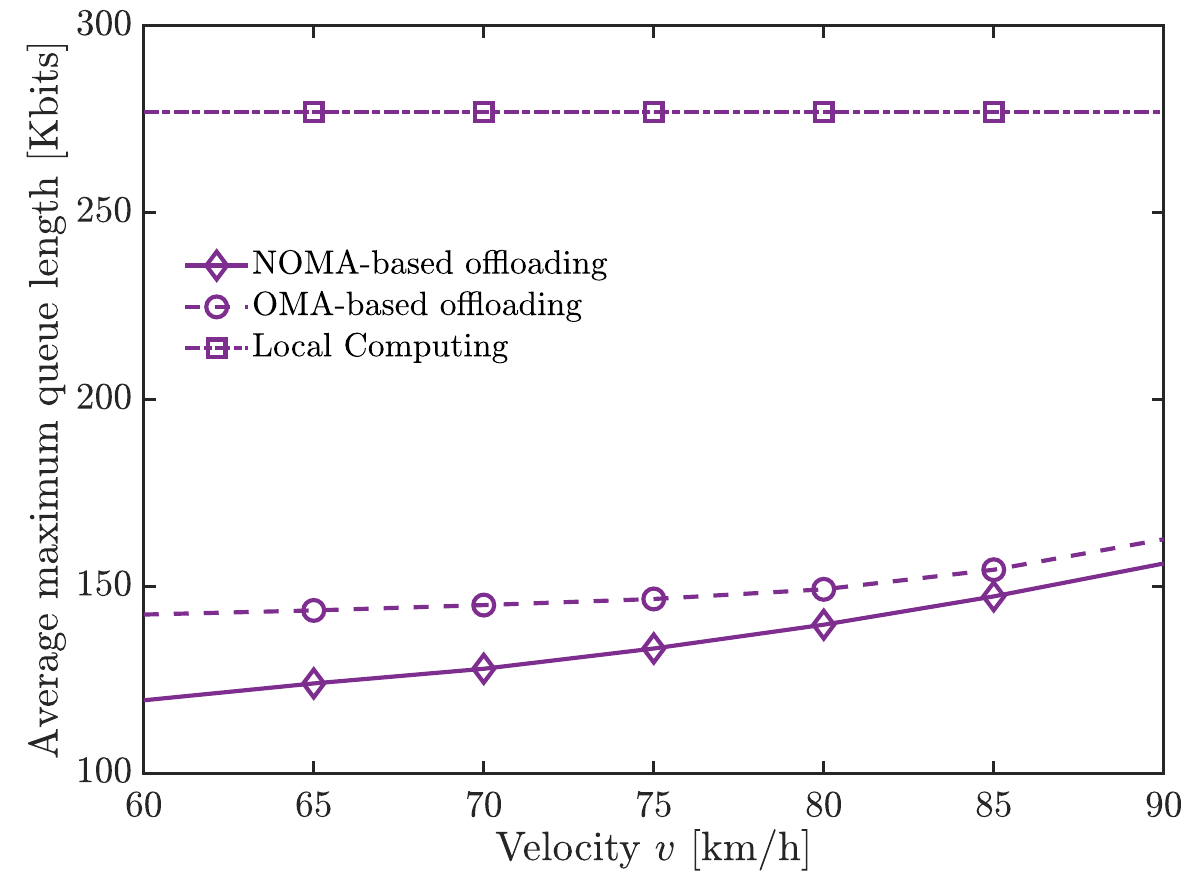}
		\caption{Average maximum queue length versus driving velocity evaluated by simulations.}
		\label{queue-velocity}
	\end{figure}

	To study the effect of driving velocity on NOMA-based computation offloading, Figure~\ref{queue-velocity} plots the average maximum queue length over 10000 TSs at vehicles versus the velocity $v$ under different offloading schemes. The weighting factor $V$ and the channel estimation error variance $\sigma_h^2$ are set to $1\times10^{-8}$ and $1\times10^{-16}$ for all velocities, respectively. Intuitively, the average queue length is expected to increase with the velocity $v$ for both NOMA-based and OMA-based offloading. This is because the spacing among vehicles becomes larger for a higher velocity, which reduces the offloading probability at the same weighting factor $V$. Moreover, the performance gap between NOMA-based and OMA-based offloading is reduced upon increasing the velocity $v$, which would tend to zero for a sufficiently large velocity $v$. Explicitly, computation offloading will seldom be used over long distances.
	
\begin{figure}[!t]
	 	\centering
	 	\includegraphics[width=0.45\textwidth]{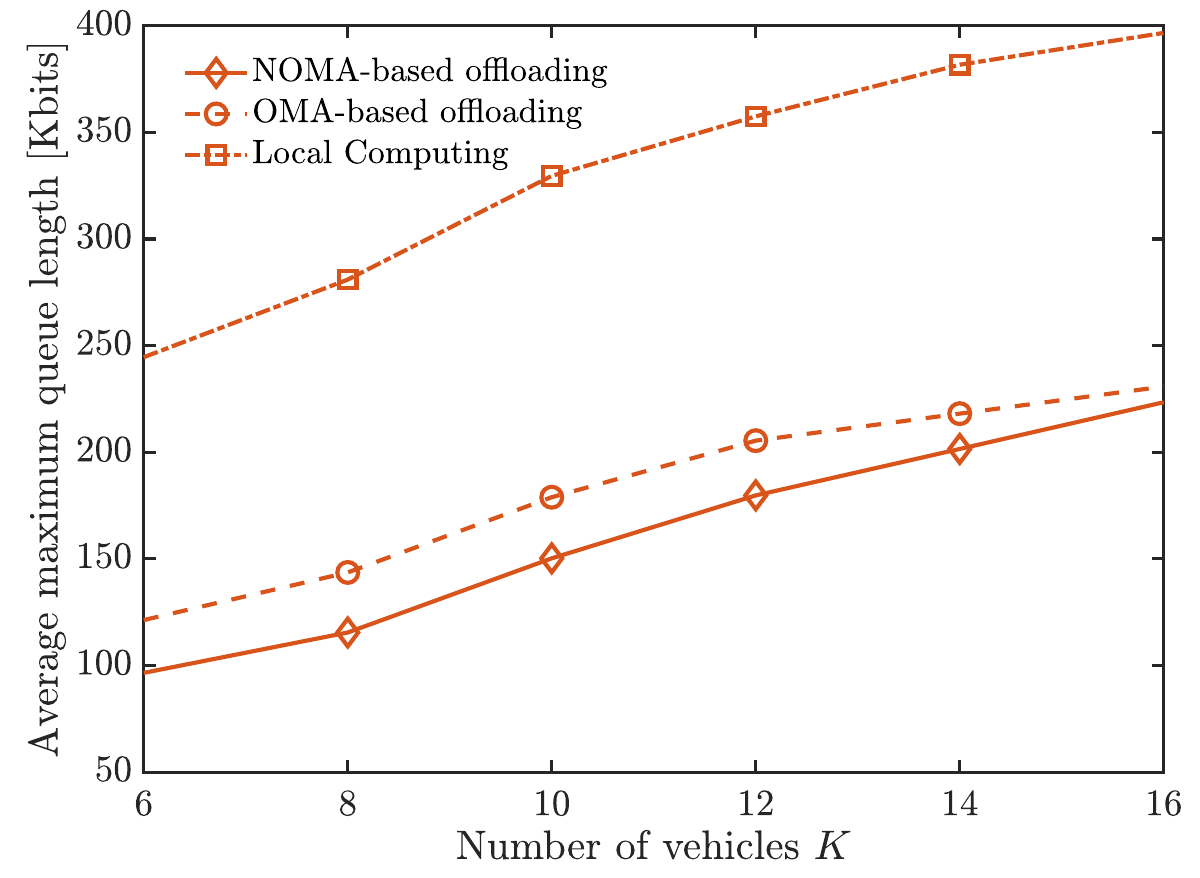}
	 	\caption{Average maximum queue length versus the number of vehicles evaluated by simulations.}
	 	\label{queue-numberK}
	 \end{figure}

	 Furthermore, the average maximum queue length at the vehicles versus  number of vehicles $K$ under different offloading schemes is plotted in Fig. \ref{queue-numberK}. We conclude from Fig. \ref{queue-numberK} that the average maximum queue length will increase, when the number of vehicles $K$ is increased. Nevertheless, compared to local computing, NOMA-based computation offloading and its OMA-based counterpart is less sensitive to the number of vehicles. Moreover, NOMA-based computation offloading always shows its superiority over the other two schemes. We note that the performance advantage of NOMA-based computation offloading over its OMA-based counterpart is beneficial, but it is gradually reduced upon increasing the total number $K$ of vehicles in a platoon. {\color{black}Intuitively, the average number of vehicles in overloaded status increases, as the cardinality $K$ of the vehicular platoon grows. Therefore, the bandwidth allocated to each vehicle having computation-intensive missions is reduced, which decreases the data offloading capability of each overloaded vehicle and hence narrows the gap between the NOMA-based offloading scheme and its OMA-based counterpart.}
	 
\begin{figure}[!t]
		\centering
		\includegraphics[width=0.45\textwidth]{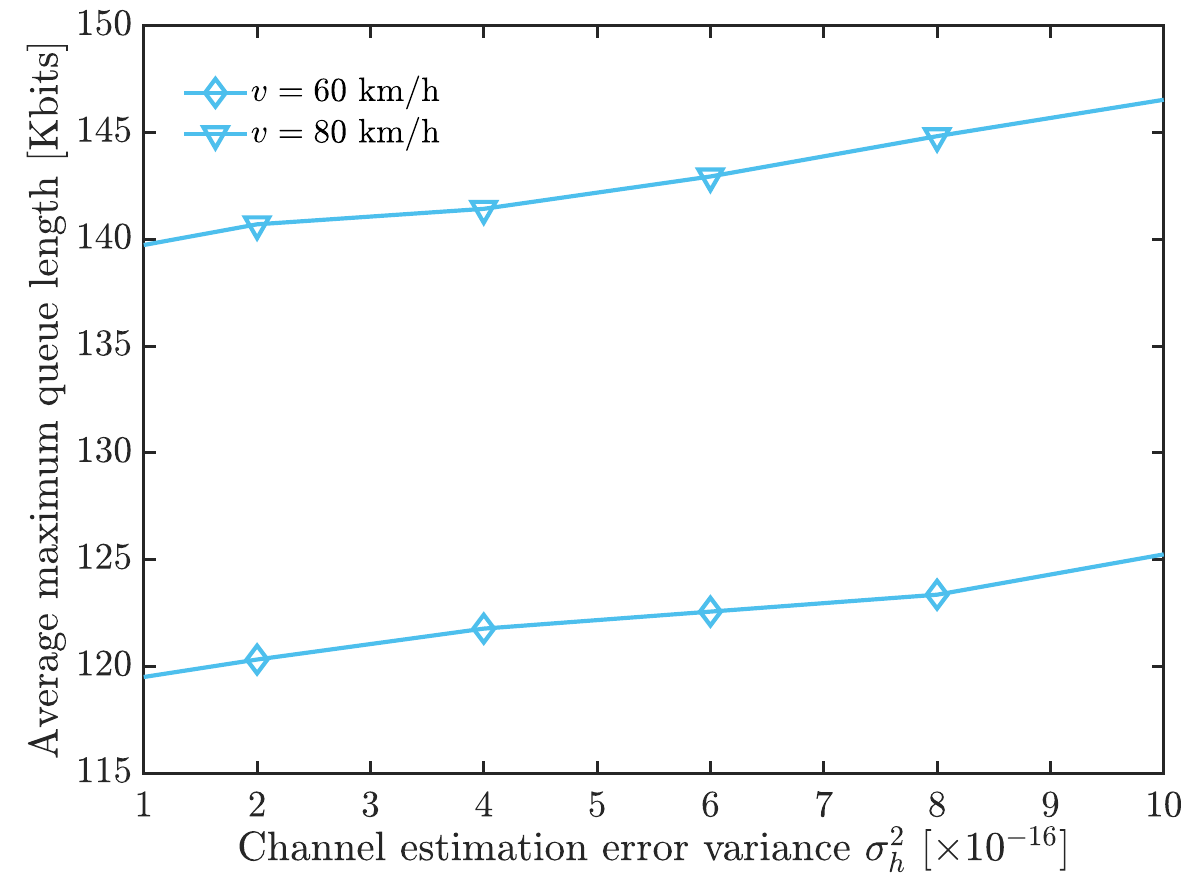}
		\caption{Average maximum queue length versus the channel estimation error variance evaluated by simulations for NOMA-based computation offloading.}
		\label{queue-variance}
	\end{figure}

 	Observe from Fig. \ref{queue-variance} that the increase of the channel estimation error variance $\sigma_h^2$ lengthens the average maximum queue length at vehicles for NOMA-based computation offloading. That is because as the channel estimation error increases, the offloading data rate among vehicles will decrease in order to restrict the outage probability of each data transmission. Meanwhile, it can be found that the slope of the curve with larger driving velocity $v$ is steeper. Explicitly, a larger value of $v$ represents a larger distance between any two adjacent vehicles, which causes a worse channel quality for data transmission. Therefore, the estimation error has a more significant impact on the average maximum queue length with the increase of driving velocity.
 	
\begin{figure}[!t]
		\centering
		\includegraphics[width=0.45\textwidth]{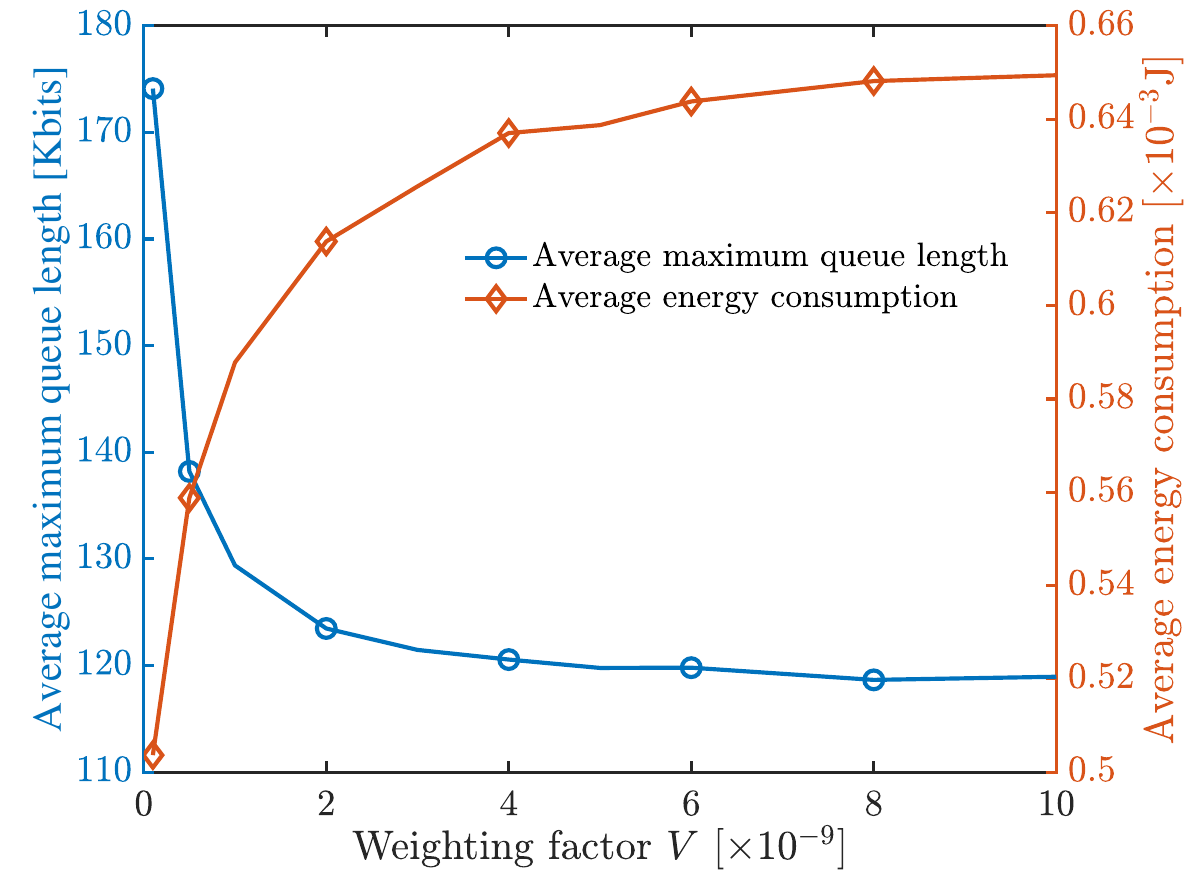}
		\caption{Average maximum queue length and average energy consumption versus the weighting factor.}
		\label{queue_V}
	\end{figure}

	{In Fig. \ref{queue_V}, we investigate the impact of the weighting factor $V$ both on the average maximum queue length and on the average energy consumption of the vehicular platoons considered. The number of vehicles $K$, the driving velocity $v$, and the channel estimation error variance $\sigma_h^2$ are set to $8$, $60$ km/h, and $1\times10^{-16}$, respectively. According to Fig. \ref{queue_V}, it can be observed that the average maximum queue length decreases, when the weighting factor $V$ is increased. In contrast to the average maximum queue length, the average energy consumption increases with the increase of the weighting factor $V$. These phenomena indicate that a flexible tradeoff between the system's delay and energy consumption can be achieved by tuning the weighting factor $V$.}
	
	\section{Conclusions}
	A dynamic NOMA-based robust computation offloading scheme was conceived and analyzed for a vehicular platoon. Then the problem of minimizing the average maximum task queue has been formulated to optimize the offloading decisions and the associated resource allocation. By converting the problem formulated into an online optimization problem, we can solve the primary problem by solely taking into account the current queue state and channel state. Then the classic SCA method has been adopted for solving the non-convex transformed problem. Our numerical results have shown that NOMA-based computation offloading outperforms both the OMA-based computation offloading and the local computing scheme. Moreover, the effect of driving velocity and the channel estimation error of the NOMA-based computation offloading scheme have been investigated. Our simulation results have revealed that a lower velocity and lower channel estimation error will shorten the task queues at the vehicles. In our future research, we may conceive a similar scheme for drones flying in formation.
	
	\appendices
	\section{Proof of Lemma 1}
	\begin{IEEEproof}
		According to (\ref{energy_queue}), we can proceed as follows:
		\begin{equation*}
			\begin{split}
				&U_{k}^2(t+1) = \left\{\left[U_k(t)+E_k^{\mathrm{total}}(t)-E_0\right]^{+}\right\}^2\\
				&\le U_k^2(t)+\left[E_k^{\mathrm{total}}(t)-E_0\right]^2+2U_k(t)\left[E_k^{\mathrm{total}}(t)-E_0\right].
			\end{split}
		\end{equation*}
		Then, the Lyapunov drift may be rewritten as
		\begin{equation*}
			\begin{split}
				&\Delta(\mathbf{\Theta} (t))=\frac{1}{2}\sum_{k \in \mathcal{K}}U_k^2(t+1)-\frac{1}{2}\sum_{k \in \mathcal{K}}U_k^2(t)\\
				&=\frac{1}{2}\sum_{k \in \mathcal{K}}\left\{U_k^2(t+1)-U_k^2(t)\right\}\\
				&\le\frac{1}{2}\sum_{k \in \mathcal{K}}\left\{ U_k^2(t)+\left[E_k^{\mathrm{total}}(t)-E_0\right]^2\right\}+\sum_{k \in \mathcal{K}} U_k(t)\left[E_k^{\mathrm{total}}(t)-E_0\right]\\
				&\le \Phi+\sum_{k \in \mathcal{K}} U_k(t)\left[E_k^{\mathrm{total}}(t)-E_0\right],
			\end{split}
		\end{equation*}
		where $\Phi$ is a finite constant denoting the upper bound of $\frac{1}{2}\sum\nolimits_{k \in \mathcal{K}}\left\{ U_k^2(t)+\left[E_k^{\mathrm{total}}(t)-E_0\right]^2\right\}$.
		
		Accordingly, the upper bound of the Lyapunov drift-plus-penalty function is formulated as
		\begin{equation}
			\begin{split}
				\Delta_V(\mathbf{\Theta} (t))&=\mathbb{E}\left[\Delta(\mathbf{\Theta} (t))+VE^{\textrm{total}}(t)|\mathbf{\Theta}(t)\right]\\
				&=\mathbb{E}\left[\Delta(\mathbf{\Theta} (t))|\mathbf{\Theta}(t)\right]+\mathbb{E}\left[VE^{\textrm{total}}(t)|\mathbf{\Theta}(t)\right]\\
				&\le \Phi+\mathbb{E}\left\{\sum_{k \in \mathcal{K}} U_k(t)\left[E_k^{\mathrm{total}}(t)-E_0\right]|\mathbf{\Theta}(t)\right\}\\
				&\quad+\mathbb{E}\left[VE^{\textrm{total}}(t)|\mathbf{\Theta}(t)\right],
			\end{split}
		\end{equation}
		which completes the proof.
	\end{IEEEproof}

	\bibliographystyle{IEEEtran}

\end{document}